\documentclass[doublecol]{epl2} 
\usepackage{graphicx,amsfonts,amsmath,amssymb,color}
\usepackage{color}
 \definecolor{darkgreen}{rgb}{0,0.6,0}
 \definecolor{orange}{rgb}{1,0.64,0.0}

\newcommand{\bi}[1]{Fig.~\ref{fig:#1}}
\newcommand{\e}[1]{Eq.~(\ref{eq:#1})}

\newcommand{\lr}[1]{\langle #1 \rangle}

\title{Noise-controlled bistability in an excitable system with positive feedback}
\shorttitle{Noise-controlled bistability in an excitable system with positive feedback} 

\author{Justus A. Kromer\inst{1} \and Reynaldo D. Pinto\inst{2} \and Benjamin Lindner\inst{1,3} \and Lutz Schimansky-Geier\inst{1,3}}
\shortauthor{J.A. Kromer \etal}

\institute{                    
  \inst{1} Department of Physics, Humboldt-Universit\"{a}t zu Berlin, Newtonstr. 15, 12489 Berlin, Germany\\
  \inst{2} Laborat\'orio de Neurodin\^amica/Neurobiof\'isica, Universidade de S\~ao Paulo, S\~ao Carlos, SP, Brazil\\
  \inst{3} Bernstein Center for Computational Neuroscience Berlin, Germany
  
}


\abstract{
We study the interplay between noise and a positive feedback mechanism in an excitable system that generates events.  We show that such a system can exhibit a bistability in the dynamics of the event generation 
(states of low and high activity).  The stability of the two states is determined by the strength of the noise  such that a change of noise intensity  permits complete control over the probabilities with which the two states are occupied. 
The bistability also has strong implications for the regularity of the event generation. While the irregularity of the interevent interval (short-time variability) and of the 
asymptotic Fano factor of the event count (long-time variability) is limited if the system is only in one of the two states, we show that both measures of variability display giant values if both states are equally likely.
The long-time variability is additionally amplified by long-range positive correlations of the interevent intervals.    
}

\begin{document}

\maketitle

\section{Introduction} 
Excitable systems that generate all-orx-none responses - or, generally speaking, events - are ubiquitous. 
Prominent examples are the action potentials of nerve cells \cite{izhikevich2007dynamical}, concentration pulses of intracellular calcium \cite{SkuKet08}, dropout events of lasing intensity in excitable lasers with optical feedback \cite{giacomelli2000experimental}, and  
blackouts of the power supply \cite{dobson2007complex}, to name but a few. Excitability can only occur in systems that are far from thermodynamic equilibrium and is often accompanied by a considerable amount of fluctuations \cite{LinGar04, carelli2005whole}. This is particularly interesting from a theoretical point of view because the interaction of noise and the inherently nonlinear mechanism of excitability leads to nontrivial effects such as the well-studied phenomena of stochastic resonance and coherence resonance \cite{PikKur97,LinGar04}.  

As with any simple model, the textbook example of an excitable system is rarely encountered in reality. Many excitable systems are subject to additional slow processes that modulate, control or feedback onto the event-generating mechanism.
Neurons, for instance, are often subject to adaptation or enhancement processes acting on multiple time scales \cite{BenHer03,pozzorini2013temporal}. Calcium oscillations (regarded as the consequence of an excitability mechanism) depend on other processes in the cell, that are in turn also influenced by the Ca concentration \cite{falcke2004reading}. Even a man-made non-equilibrium device as the laser can display an excitability that does not conform with a simple low-dimensional dynamical model but should be rather looked upon as an excitable  subsystem  embedded in a much higher dimensional system \cite{PhysRevLett.76.220}. So far, most researchers have focussed on the effect of a negative (down-regulating) feedback. However, excitable systems can be also subject to 
positive feedback loops. 
Neurons, for instance,  can be self-excitatory via so-called autapses \cite{saada2009autaptic,bekkers2009synaptic} or by means of the external potassium concentration \cite{postnov2009dynamical,frohlich2006slow}. Positive feedback is also present in genetic circuits, e.g. in the lactose utilization network of \textit{Escherichia coli} \cite{ozbudak2004multistability}.  

From a theoretical point of view, positive feedback is interesting because, as we will show in this study, it can give raise to bistability in the event-generating rate, comparable
to the bistability between low and high-activity states recently observed in neural networks \cite{grau2012noise,mikkelsen2013emergence}. 

This bistability is controlled by the fluctuations in a novel way and also influences remarkably the regularity of the event generation.

\section{Noisy oscillator with event-triggered feedback}  Our starting point is a paradigmatic excitable system,  Adler's equation for the phase
evolution driven by a white Gaussian noise $\xi(t)$
\begin{align}
\label{eq:CompleteSystemDet}
 \dot{\phi}(t) = \omega_{0} - \sin[\phi (t)]+ \sqrt{2 D} \xi(t),
\end{align} 
where $D$ denotes the noise intensity and  $\langle \xi(t) \xi(t') \rangle = \delta(t-t')$; here and in the following $\langle f(t) \rangle$ denotes the time average of the function $f(t)$. Whenever the phase crosses $2\pi$, we postulate the occurrence of an event, register its event time $t_i$, and reset the phase to zero ($\phi \rightarrow 0$). We can define an event train, with delta peaks at the event times:
\begin{equation}
\label{eq:spiketrain}
  x(t)=\sum_{i} \delta(t-t_{i}).
\end{equation}
Important statistics of the series of events are (i) those of the interevent intervals (IEI) \mbox{$\Delta t_i=t_{i+1}-t_i$}, that characterize the regularity of events on a short time scale and (ii) those of the event count $N(T)=\int_0^T dt' x(t')$ that describes for large time window $T$ the long-term variability.

We focus on a positive driving $0<\omega_0<1$ that pushes the system in the excitable regime, in which a stable node and unstable node coexist. In the deterministic system ($D=0$), a bifurcation from quiescence (no events) to a tonic regime (generation of a strictly periodic sequence of events)  occurs  at  $\omega_0=1$. 


\e{CompleteSystemDet} does not include a feedback yet. We extend the dynamics  as follows \cite{Kromer2014EventFeedback}
\begin{align}
\label{eq:CompleteSystem}
 \dot{\phi}(t) = \omega_{0}+\Delta \omega(t) - \sin[\phi (t)] + \sqrt{2 D} \xi(t).
\end{align} 
The feedback variable  $\Delta \omega(t)$ is governed by 
\begin{align}
\label{eq:TriggeredFeedback}
\tau \frac{d}{dt}\Delta \omega(t) = - \Delta \omega(t)+ 2 \pi a \ x(t).
\end{align}
 Note that $\Delta \omega(t)$ is solely driven by the event train 
\e{spiketrain}  generated by  the excitable system. This kind of event-triggered feedback is particularly simple but also mimics the way different excitable systems
are subjected to feedback.
 
\begin{figure}
\onefigure[width=0.7\linewidth]{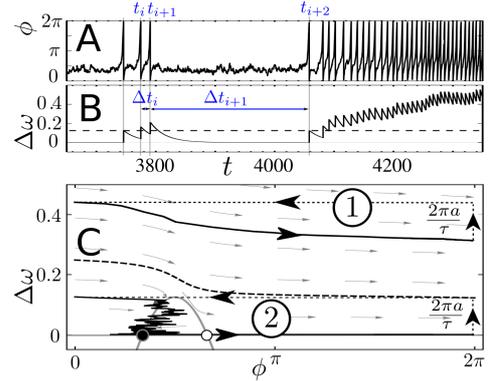}
\caption{Phase $\phi$ (A) and feedback $\Delta \omega$ (B) as functions of time and representative trajectories during one IEI for the 
high-activity $\textcircled{1}$ and the low-activity state $\textcircled{2}$ in the $(\phi, \Delta \omega)$ space (C).
For large $\tau$, both states are separated by an unstable limit cycle (dashed thick curve, in C), which becomes an homoclinic orbit
for small $\tau$ (see also Ref. \cite{Kromer2014EventFeedback}).
B: Labels illustrate the event times $t_i$ and IEIs $\Delta t_i$.
The dashed line illustrates $\Delta \omega$ upon reset for the unstable limit cycle. 
C: The reset condition (Dashed arrows), the nullclines (gray curves), and the deterministic velocity field (gray arrows) are illustrated. Parameters: $a=0.5$, $\omega_0=0.875$, $\tau=25$, and $D=0.02$.}
\label{fig:trajectories}
\end{figure}

The parameter $a$ scales the kick strength; here we consider exclusively positive 
feedback ($a>0$), for which  a kick effectively increases the driving. 
Finally, $\tau$ is the timescale of the exponential decay and is assumed to be large
(slow feedback). Sample traces of $\phi(t)$ and $\Delta \omega(t)$ and trajectories in the phase plane $ (\phi,\Delta \omega)$ are shown in Fig. \ref{fig:trajectories}.

\section{Bistability of the event generation} 
As demonstrated in Ref. \cite{Kromer2014EventFeedback},  a strong positive feedback in the deterministic system ($D=0$) leads to the coexistence of two asymptotic stable solutions. In dependence on the initial condition either a fixed point or a limit cycle is approached. Here we explore the effect of fluctuations in this bistable regime.

With noise the system is able to escape from the fixed point due to noisy excitations, which leads to event generation at a low rate. In this low-activity state (indicated by $\textcircled{2}$ in \bi{trajectories}C), $\Delta \omega(t)$ is close to zero and IEIs are approximately exponentially  distributed (Poisson statistics). In contrast, the high-activity state (marked by $\textcircled{1}$ in \bi{trajectories}C), corresponding to the limit-cycle solution of the deterministic system, shows a large average value of
$\Delta \omega(t)$ and exhibits a rather regular event generation.

Transitions between the two states of event generation are possible due to noise. 
Interestingly, there is an asymmetry in the transitions for $\tau>2 \pi a/(1-\omega_0)$.  
Escapes from the low-activity to the high-activity state are only possible within 
a sequence of short IEIs ($\Delta t_i < \tau$) as  is also visible in \bi{trajectories} (A and B). 
In such a sequence, the feedback can build up and, when passing the unstable limit cycle, stabilize the system in a tonic regime of high activity.
In contrast, the high-activity state can be left again within one very long interval ($\Delta t_i \gg \tau$).

\section{Theoretical calculation of bistable states} We assume a slow feedback dynamics, i.e. $\tau \gg \langle \Delta t_i \rangle$ for which  
the feedback variable is hardly affected by fluctuations of individual IEIs. 
In this case, $\Delta \omega$ can be approximated by an average feedback strength
\cite{Kromer2014EventFeedback}
\begin{align}
\Delta \omega(t) \approx  \lr{\Delta \omega} = 2 \pi a\lr{x(t)}.
\end{align}
The last equality follows from time-averaging \e{TriggeredFeedback}. Using this expression in \e{CompleteSystem} leads to a one-dimensional problem, in which, however, 
the effective driving depends on the system's output statistics (i.e. the event rate $\lr{x(t)}$) in a self-consistent mean-field-like manner. The one-dimensional system is equivalent to the Brownian motion in a tilted periodic potential $U(\phi; \lr{\Delta \omega}) = -(\omega_0 + \langle \Delta \omega \rangle) \phi - \cos(\phi)$, which depends on $\lr{\Delta \omega}$.

\begin{figure}[t]
    \onefigure[width=0.8\linewidth]{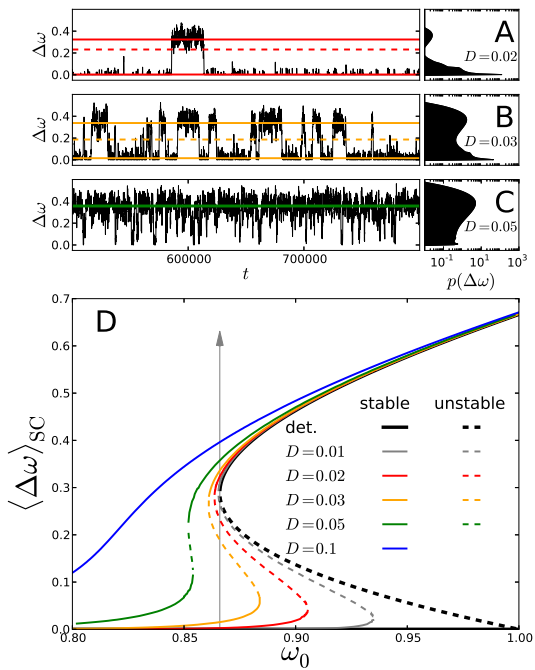}
\caption{Solutions of Eq. (\ref{eq:mFPT}) for different $D$ and of Eq. (\ref{eq:mFPTweakNOISe}) (det.) (D). For $\omega_0=\sqrt{1-a^2}$ (see arrow)
trajectories\protect\footnotemark[4] for different noise intensities and corresponding distribution of $\Delta \omega$\protect\footnotemark[5] are shown in the plots (A,B,C). Horizontal lines indicate the corresponding 
stable (straight) and unstable (dashed) solution for $\lr{\Delta \omega}_{\rm SC}$ obtained from Eq. (\ref{eq:mFPT}). Parameters: $a=0.5$; for (A,B,C) we used $\tau=50$.}
 \label{fig:Rate}
\end{figure} 
\footnotetext[4]{We used a stochastic Euler method (SEM) with dynamic time step $\Delta t = \text{min}[2 \pi/(100|v_{\text{drift}}|),0.1]$. Here $v_{\text{drift}}$ 
is the deterministic part of Eq. (\ref{eq:CompleteSystem}). $0.2\%$ of stored $\Delta \omega$ were plotted.} 
\footnotetext[5]{$p(\Delta \omega)$ was obtained by integrating the stationary distribution $p(\phi, \Delta \omega)$ over $\phi$. $p(\phi, \Delta \omega)$ was calculated with the rare-event algorithm \cite{kromer2013weighted}.
Parameters according to Ref. \cite{kromer2013weighted}:
    $N=2$, $h=0.1$, $L_{\phi}^{-}=-1.07$, $L_{\phi}^{+}=2 \pi$, $L_{\Delta \omega}^{-}=-0.05$, $L_{\Delta \omega}^{+}=1.5$, $\Delta \phi_{box}=0.01 \sqrt{2 D h}$, $\Delta \omega_{box}=0.5 h/\tau$,
    $n_1=1$, $n_2=5$, $N_T=10000$, and $N_{T0}=T_{\text{therm}}/(n_1 n_2 h)$. 
    The equilibration times $T_{\text{therm}}$ were set to:
      $T_{\text{therm}}=10000$ ($D=0.05$), 
      $T_{\text{therm}}=30000$ ($D=0.03$), and 
      $T_{\text{therm}}=200000$ ($D=0.02$).}
The self-consistent mean value $\lr{\Delta \omega}_{\rm SC}$ is obtained from requiring that for a given $\lr{\Delta \omega}$, the event-generating rate corresponding to the mean velocity of the Brownian particle in an inclined periodic potential \cite{stratonovich1967} yields the very same $\lr{\Delta \omega}$:
\begin{align}
\label{eq:mFPT}
\lr{\Delta \omega}_{\rm SC}=2 \pi a r_{\rm SC}=
\frac{a D \sinh(\frac{\pi}{D}(\omega_0+\lr{\Delta \omega}_{\rm SC}))}{\pi |I_{i\frac{\omega_0+\lr{\Delta \omega}_{\rm SC}}{D}}(\frac{1}{D})|^2}
\end{align}
Here $I_z(y)$ denotes the modified Bessel function of the first kind and imaginary order $z$, $i$ the imaginary unit, and $r_{\rm SC}$ the rate of event generation associated with $\lr{\Delta \omega}_{\rm SC}$.
Graphical solutions of this equation are illustrated in \mbox{Fig. \ref{fig:Rate}}D. 
They were obtained by searching for intersections of the left-hand and the right-hand side. 
Remarkably, there are up to three solutions, two of which correspond to stable (solid lines) and one of which to an unstable (dashed lines) asymptotic states. 

Analytical approximations for the mean IEI in the deterministic limit ($D =0$) and for \mbox{$\tau \gg \langle \Delta t_i \rangle$} were previously derived in Ref. \cite{Kromer2014EventFeedback}. 
Here, up to two non-trivial solutions exist, which result in two solutions for $\lr{\Delta \omega}_{\rm SC}$:
\begin{align}
\label{eq:mFPTweakNOISe}
\lr{\Delta \omega}_{\rm SC,1/2}=2 \pi a r_{\rm SC, 1/2}=\frac{a (1-\omega_0^2)}{\pm \sqrt{\omega_0^2+(a^2-1)}+a \omega_0}.
\end{align}
The trivial solution $\lr{\Delta \omega}_{\rm SC}=0$ exists for all $\omega_0\le 1$. Both, graphical solutions for $D > 0$ and analytical approximations for $D=0$ are depicted in \mbox{Fig. \ref{fig:Rate}}D.


Interestingly, the number of self-consistent solutions is controlled by the noise intensity. Only for a moderate range of noise intensities do two stable solutions coexist. For strong fluctuations, only the limit cycle solution survives.

\section{Noise-controlled bistability} 
In order to study the role of noise in more detail, we fix the parameters of the deterministic system and study trajectories for different noise intensities.
Results are illustrated in Fig. \ref{fig:Rate} (A,B,C).
For a weak noise (A), the system stays most of the time in the low-activity state. The corresponding distribution of the feedback variable $\Delta \omega$ possesses a well-pronounced maximum at the stable low-feedback solution of Eq. (\ref{eq:mFPT}). A second minor maximum at the stable high-feedback solution indicates that the high-activity state exist but is occupied only rarely. Increasing solely the strength of fluctuations stabilizes the high-activity state (B). Here $p(\Delta \omega)$
shows a well-pronounced bimodality with two peaks at solutions of Eq. (\ref{eq:mFPT}); probabilities in both states (integral over the respective peak) are close to each other.  Finally, for a strong noise (C), only the high-activity state remains stable. Here excursions to low feedback strengths immediately relax to the single high-feedback solution of Eq. (\ref{eq:mFPT}) and the system resembles a noisy excitable system.

The above-described dependence of the occupation probability on the noise strength differs radically from what is seen in a classical bistable system with noise.  Consider as an example  a Brownian particle in an asymmetric bistable potential with two minima at $x_-$ and $x_+$ $(x_-<x_+)$ of different depths ($U(x_-)<U(x_+)$)
and a maximum at $x_0$ in between. Such a particle would be  almost certainly in the deeper well if noise is very weak 
($p_-=\int_{-\infty}^{x_0} P(x) \simeq 1$) and it could be encountered with equal probability in the domains of attraction of both minima if noise is very strong ($p_-\simeq 1/2$). However, there is no way to choose a noise such that the particle prefers the domain of $x_+$ and thus by noise we can only change $p_- \in [0.5,1]$ and $p_+\in [0,0.5]$.  In marked contrast to this, in our system, 
 the occupation probabilities $p_\pm$ of the two states can be tuned by the noise in the full range $[0,1]$.




Noise-controlled bistability is most pronounced for $\omega_0$  close to but smaller than $ \sqrt{1-a^2}$ (see Fig. \ref{fig:Rate}D), which is the border of coexisting solutions in the deterministic limit Eq. (\ref{eq:mFPTweakNOISe}). It arises from the 
fact that the mean firing rate of the original excitable system increases monotonically with the noise 
intensity \cite{stratonovich1967}. This increases the average feedback strength and drives the system closer to the tonic regime.
\\
\section{Control of variability of the event generation} 
The noise term does not only control the coexistence of states but also the regularity of event generation in each of the states. 
Here we analyze this regularity on the short time scale of a single IEI in terms of its coefficient of variation (CV) and on long time scales by means of the asymptotic Fano factor of the number of events. 

\begin{figure}[t]
\onefigure[width=0.8\linewidth]{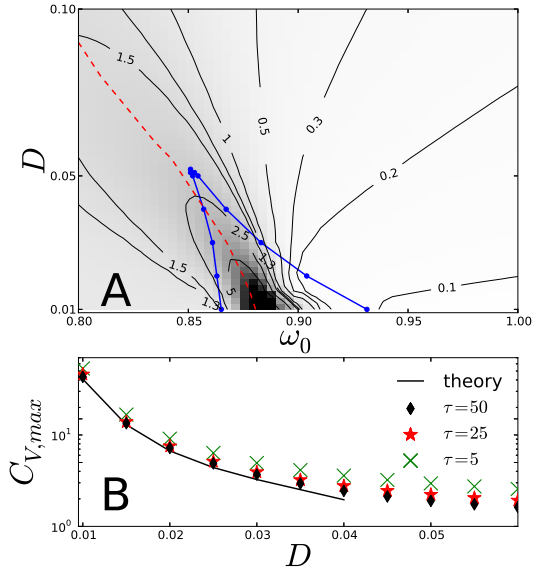}
\caption{CV for different $D$ and $\omega_0$ (A), and its maximum value for fixed $D$ (B)\protect\footnotemark[6]. 
 A: Lines show the $C_V$ levels (black), the maximum CV for a fixed $D$ (red, dashed), and border the region of multiple solutions in Eq. (\ref{eq:mFPT}) (blue).
Dark and light regions illustrate high and low CVs, respectively.  
B: $C_{V,max}$ obtained from Eq. (\ref{eq:twoStateApprox}) (theory) and results from simulations (symbols) for different $\tau$. 
Parameters: $a=0.5$, $\tau=25$ (A).}
\label{fig:CVResults}
\end{figure}
\footnotetext[6]{Trajectories of $10^7$ IEIs were simulated using SEM with $\Delta t=2 \pi \text{min}[1/10 ,1/(10000|v_{\text{drift}}|)]$ 
and $\Delta t=2 \pi \text{min}[1/10 ,1/(100|v_{\text{drift}}|)]$ for $D \leq 0.02$ and $\omega_0 \leq 0.85$.
For (B) $10^8$ IEIs were simulated for $\tau=50$ and $D\leq 0.02$.}

\subsection{IEI variability} 
The variability of individual IEIs is  quantified by the CV
\begin{align}
\label{eqs:CV}
 C_V=\frac{\sqrt{\langle (\Delta t_i - \langle \Delta t_i \rangle)^2 \rangle}}{\langle \Delta t_i \rangle}.
\end{align}
For comparison: a perfectly periodic event train possesses $C_V=0$, a rare-event (Poisson-like) generation  corresponds to $C_V=1$, and burst-like event generation is characterized by $C_V>1$. 

Fig.~\ref{fig:CVResults}A shows numerical results for the CV and reveals 
a nontrivial dependence of this regularity measure on
the noise intensity $D$ and the driving $\omega_0$.
First of all, there are simple limit cases for weak noise. For weak driving (e.g. $\omega_0=0.8$), there is only the low-activity state occupied and $C_V\to 1$ for $D\to0 $; for $\omega_0$ approaching the bifurcation point ($\omega_0=1$), the high-activity state is the only stable state and thus $C_V \to 0$ for $D\to0 $. Interestingly, we find a very high IEI variability in between these limits, i.e. in the region of multiple solutions for $\lr{\Delta \omega}_{\rm SC}$.

In order to investigate the increase of IEI variability in the regime of coexisting states analytically, we 
follow the approach of Ref. \cite{schwalger2012interspike} and approximate the system by a two-state Markov Process. We assume the system to perform transitions between the
high-activity ($H$) and the low-activity ($L$) state with constant rates $\lambda^{\text{H}}$ (for the transition $H \rightarrow L$) and 
$\lambda^{\text{L}}$ for the opposite transition. Mean waiting times in the respective state are then $T^{\text{H,L}}=1/\lambda^{\text{H,L}}$.
 Additionally, each state is characterized by its individual rate for the event generation $r^{\text{H}}=1/\langle \Delta t_i^{\text{H}} \rangle$ and $r^{\text{L}}=1/\langle \Delta t_i^{\text{L}} \rangle$ and its individual CV, $C_{V,\text{H}}$ and $C_{V,\text{L}}$. Furthermore, to simplify the notation, we introduce 
the ratio of waiting times $\alpha$ and the ratio of the rates for event generation in the respective states $\gamma$:
\begin{align}
\label{eqs:alphaGamma}
\alpha:=T^{\text{H}}/T^{\text{L}} \ \ \ \ \gamma:=r^{\text{H}} /r^{\text{L}}.
\end{align}

Neglecting the influence of transitions between states on the statistics (instantaneous switching), we can apply the results of Ref. \cite{schwalger2012interspike} for the moments of the IEI density, which in general read
\begin{align}
\label{eqs:moments}
 \langle (\Delta t_i)^n \rangle = p_{st}^{\text{H}} \langle (\Delta t_i^{\text{H}})^n \rangle +  p_{st}^{\text{L}} \langle (\Delta t_i^{\text{L}})^n \rangle.
\end{align}
Here $p_{st}^{\text{H}} = \alpha \gamma /(1+ \alpha \gamma)$ and $p_{st}^{\text{L}} = 1-p_{st}^{\text{H}}$ denote the stationary probability to draw an IEI from either state.

Using Eqs. (\ref{eqs:CV}), (\ref{eqs:alphaGamma}), and (\ref{eqs:moments}) we obtain for the CV
\begin{align}
\label{eq:CV}
 C_V = \sqrt{ \frac{1+\alpha \gamma}{(1+\alpha)^2}\left[\frac{\alpha}{\gamma}(C_{V,\text{H}}^2+1)+(C_{V,\text{L}}^2+1)\right]-1}.
\end{align}
Solely varying $\alpha$ yields $C_{V}=C_{V,L}$ for $\alpha \rightarrow 0$ (pure excitable) and
$C_{V}=C_{V,H}$ for $\alpha \rightarrow \infty$ (pure tonic behavior). 
Because the rate of event generation is much higher in the high-activity state, we can assume $\gamma \gg 1$. 
For such $\gamma$, the CV goes through a maximum $C_V=C_{V,max}$ at some $\alpha_{max}$. 
This suggests that the pronounced maximum of the $C_V$ in Fig. \ref{fig:CVResults}A is mainly caused by the effect of $\omega_0$ and $D$ on the parameter $\alpha$, the states' occupation probabilities. 

From Eq. (\ref{eq:CV}) one can calculate $\alpha_{max}$ for a given $\gamma$.
One obtains
\begin{align}
\label{eq:alpha}
 \alpha_{max} \approx 1+ \frac{2}{\gamma} \frac{C_{V,\text{H}}^2-C_{V,\text{L}}^2}{1+C_{V,\text{L}}^2}
\end{align}
for the first order in $1/\gamma$. Because the second term is only a small correction, 
we find that the CV is maximized if the occupation probability of low and high-activity states are close to each other (corresponding to $\alpha \approx 1$). 

Using Eqs. (\ref{eq:CV}) and (\ref{eq:alpha}) we find for the maximal CV 
\begin{align}
\begin{aligned}
\label{eq:twoStateApprox}
 C_{V,max} &= \frac{1}{2}\sqrt{\gamma(1+C_{V,\text{L}}^2)}.\\ 
\end{aligned}
\end{align}
Clearly, $C_{V,max}$ diverges for $\gamma \rightarrow \infty$. This corresponds to the limit of weak noise, for which we find that \mbox{$C_{V,\text{H}}\rightarrow 0$}, \mbox{$C_{V,\text{L}}\rightarrow 1$}, \mbox{$r^{\text{H}} \rightarrow \text{const.}$} [Eq. (\ref{eq:mFPTweakNOISe})], and \mbox{$r^{\text{L}} \rightarrow 0$} because the occurrence of events disappears in the excitable regime for vanishing noise.


In order to test Eq. (\ref{eq:twoStateApprox}), we proceed as follows. 
We performed numerical simulations for a fixed value of $D$ and different values of $\omega_0$ and determined with respect to $\omega_0$ the maximal value $C_{V,max}(D)$ 
attained at $\omega_{0,max}$. We can obtain an analytical estimate of $C_{V,max}(D)$ by making the following approximations in  Eq. (\ref{eq:twoStateApprox}): (i) $C_{V,\text{L}}=1$ (Poissonian firing in the low-activity state); (ii) the two event-generating rates (needed to calculate their ratio $\gamma$) are taken from the numerical solutions of Eq.~(\ref{eq:mFPT}) at the value of $D$, that we prescribed, and at 
$\omega_0=\omega_{0,max}$, that we obtained in our simulations for $\tau=50$. Both theory and simulation results can then be plotted as functions of the noise intensity $D$ (Fig. \ref{fig:CVResults}B). 

The theory approximates $C_{V,max}$ well for large $\tau \gg 1/r^{\text{H}}$ and a weak noise. It underestimates 
the simulation results for smaller $\tau$ because this  extends the region of bistability  towards smaller $\omega_0$ and 
causes a higher rate of event generation in the high-activity state 
\cite{Kromer2014EventFeedback} resulting in larger values of $\gamma$
and, therefore, in larger CVs.
Note that $\gamma$ cannot be calculated
if the maximum CV is attained outside the region of 
multiple solution of Eq. (\ref{eq:mFPT}), which is why we cannot plot 
a theoretical curve for higher noise levels ($D>0.04$) in Fig. \ref{fig:CVResults}B.
\subsection{Count variability} In order to measure the variability on long time intervals, we consider the Fano factor
$F(T)=\langle (N(T)-\langle N(T) \rangle)^2 \rangle / \langle N(T) \rangle$.
Here $N(T)$ is the number of events in the time interval $T$. Averages are taken over an ensemble of time intervals of length $T$.
We are interested in the long-time limit $F_{\infty}$ of the Fano factor, which can be related to the CV and to the serial 
correlation coefficients $\rho_n$ of IEIs at lag $n$
\begin{align}
 \rho_n=\frac{\langle (\Delta t_i - \langle \Delta t_i \rangle)(\Delta t_{i+n} - \langle \Delta t_i \rangle) \rangle}{\langle (\Delta t_i - \langle \Delta t_i \rangle)^2 \rangle}
\end{align}
by \cite{CoxLew66}
\begin{align}
\label{eq:longtimeFano}
 F_{\infty}=\lim \limits_{T \rightarrow \infty} F(T)= C_V^2 (1+2 \sum \limits_{n=1}^{\infty} \rho_n).
\end{align}

\begin{figure}[t]
\onefigure[width=0.83\linewidth]{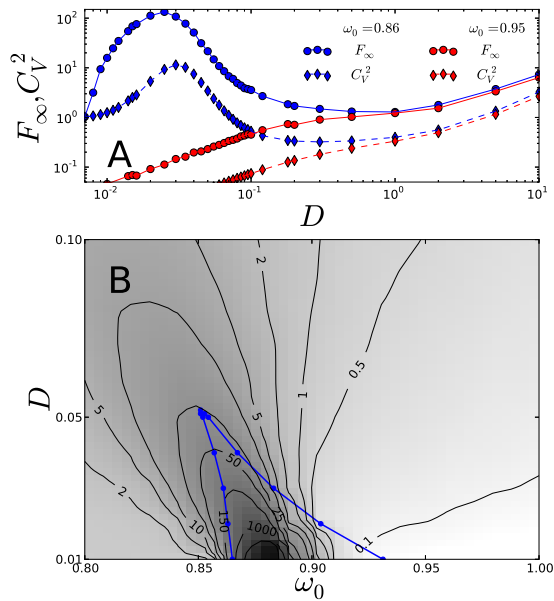}
\caption{$F_{\infty}$ and $C_V^2$ obtained from simulations for a fixed $\omega_0$ as a function of noise strength (A) and $F_{\infty}$ as a contour plot for different $\omega_0$ and $D$ (B)\protect\footnotemark[7] .
High gray levels indicate high values of $F_{\infty}$. The blue border the region of multiple solutions of Eq. (\ref{eq:mFPT}).
Parameters: $a=0.5$, $\tau=25$.}
 \label{figFanoFactor}
\end{figure}

\footnotetext[7]{Simulation details as in Fig. \ref{fig:CVResults} except that we used $\Delta t=2 \pi \text{min}[1/10 ,1/(100000|v_{\text{drift}}|)]$ for $D\geq 1$
  and $\Delta t=2 \pi \text{min}[1/10 ,1/(100|v_{\text{drift}}|)]$ also 
  for $0.85 < \omega_0 < 0.9$. For $\omega_0=0.88$ and $D=0.01$ $10^9$ IEIs were needed.
  Fano factors were calculated from ensembles of time intervals that were obtained by dividing the simulated time $T_{\text{Max}}$ in intervals of
  lengths $T_i=(T_{\text{Max}}/100)^{j/50}$, with $j=1,2,...,50$. $F_{\infty}$ was calculated by averaging the Fano factors $F(T_j)$ for $j=30,31,...50$.}

Results from simulations for $F_{\infty}$ are shown in Fig. \ref{figFanoFactor} for fixed $\omega_0$ as a function of the noise intensity (A)
and for sets of $\omega_0$ and $D$ (B).
In the weak-noise limit, $F_{\infty}$ approaches, similar to the CV, the limits $F_{\infty}\rightarrow 1$ (Poisson-like) and 
$F_{\infty}\rightarrow 0$ (pure tonic) in the respective regions. Interestingly, as observed for the CV before, $F_{\infty}$ approaches giant values in the bistable region, somewhat reminiscent of the giant diffusion observed in systems with bistable velocity dynamics \cite{LinNic08}.  
However, comparing the absolute values of $F_{\infty}$ with those obtained for $C_v^2$ (see Fig. \ref{figFanoFactor}A), we find from Eq. (\ref{eq:longtimeFano}) that IEI correlations strongly contribute to $F_{\infty}$. 
Such correlations have been found to be present over hundreds of lags in two-state systems 
with appropriate occupation probabilities \cite{schwalger2012interspike}.


Our investigation of the IEI variability suggests that the shape of the CV in the ($\omega_0$, $D$) plane is due to the modulation of the states occupation probabilities by these parameters. The low-activity state dominates for small $\omega_0$ and $D$ values ($C_v=1$), 
whereas the high-activity state dominates for high values of $\omega_0$ and/or $D$ (regular IEI statistics, $C_v<1$). Both regimes
are separated by a region of coexisting states with high variability ($C_v >1$).
Because noise affects the occupation probability, it leads to a maximum in the CV as a function of noise intensity, known as anticoherence resonance \cite{PhysRevE.66.045105}.
The maximum CV is mainly determined by the ratio of event generation rates
in both states and diverges in the weak-noise limit.

Interestingly, not only huge CVs can be achieved but also strong correlations in the system. Both in combination, however, cause giant values of the asymptotic Fano factor and can cause 
a local maximum in $F_{\infty}$ as a function of noise intensity as can be observed in Fig. \ref{figFanoFactor}A.


%

\section{Summary and conclusion}
The presence of strong positive event-triggered feedback and noise in the excitable system leads to the coexistence of a low-activity and a high-activity state.
Both states possess different individual statistics. The low-activity state statistics is mainly determined by the Poisson-like 
behavior of the original excitable system. In contrast, the high-activity state yields more regular tonic event generation. 

Because noise increases the rate of event generation in the low-activity state, stronger fluctuations cause more events and thereby stronger event-triggered feedback.
This leads to novel noise-induced phenomena.

Most importantly, external fluctuations control the occupation probabilities of the two well-distinguishable states such
that by changing the noise intensity these probabilities can be changed between zero and one. A necessary ingredient for this is that the system 
must be close to (but not within) a dynamical regime, in which two spiking states coexist. As a consequence, if 
noise is weak, only the low-activity state exists, whereas in the strong-noise regime the system resides exclusively in the high-activity state. For intermediate noise intensities both states coexists. In the limit of slow feedback, we derived an analytical theory which predicts this phenomenon  in form of a bistability  of the rate of event generation (and of the corresponding average feedback strength) for intermediate noise strength. 

Studying the IEI and count variability, we found both to reach giant values in the regime of coexisting states. The CV, which characterizes the variability of event generation at the short time scale of a single interval, can be made  large 
by increasing the ratio of event-generating rates in the high- and low-activity states. One can expect that such a large irregularity of the single interval will result in a likewise large variability at long time-scales as quantified by the long-time limit of the Fano factor. 
This statistics, however, showed an even larger variability, caused by strong positive IEI correlations that arise due to the coexistence of the two states.   


Combining  noise-controlled bistability with the fact that variability in a two-state system is maximized for equal occupation of both states, yields
a new mechanism for anticoherence resonance in excitable systems with a positive feedback mechanism. We demonstrated that anticoherence resonance 
can be observed in the variability of individual IEIs (the coefficient of variation) and in the long-time limit of the  count variability (the Fano factor) as a function of noise intensity.

Our results illustrate that bistability in event generation can result from a subtle interplay between positive feedback and fluctuations in an excitable system. 
They will be useful in interpreting observations of bistable behavior in complex systems such as up-and-down states in neural networks \cite{HolTso06,hidalgo2012stochastic} 
and the role of positive feedback, for instance due to autapses \cite{bekkers2009synaptic}. 
    
\acknowledgments
This paper was developed within the scope of IRTG 1740/TRP 2011/50151-0, funded by the DFG / FAPESP and by the BMBF (FKZ: 01GQ1001A).

\bibliographystyle{eplbib}

\end{document}